\documentclass[aps, pre, tightenlines, amsmath,amssymb]{revtex4}
\newif\ifpdf\ifx\pdfoutput\undefined\pdffalse\else\pdfoutput=1\pdftrue\fi

\usepackage{graphicx}
\usepackage{bm}
\usepackage{epsfig}

\begin{document}

\title{\bf Phase equilibria and fractionation in a polydisperse fluid}
\author{Nigel B. Wilding}
\affiliation{Department of Physics, University of Bath, Bath BA2 7AY, U.K.}
\author{Peter Sollich}
\affiliation{Department of Mathematics, King's College London, Strand, London WC2R 2LS, U.K.}

\begin{abstract}

We describe how Monte Carlo simulation within the grand canonical
ensemble can be applied to the study of phase behaviour in polydisperse
fluids. Attention is focused on the case of fixed polydispersity in
which the form of the `parent' density distribution
$\rho^\circ(\sigma)$ of the polydisperse attribute $\sigma$ is
prescribed. Recently proposed computational methods facilitate
determination of the chemical potential distribution conjugate to
$\rho^\circ(\sigma)$. By additionally incorporating extended sampling
techniques within this approach, the compositions of coexisting
(`daughter') phases can be obtained and fractionation effects
quantified.  As a case study, we investigate the liquid-vapor phase
equilibria of a size-disperse Lennard-Jones fluid exhibiting a large
($\delta=40\%$) degree of polydispersity. Cloud and shadow curves are
obtained, the latter of which exhibit a high degree of fractionation
with respect to the parent. Additionally, we observe considerable
broadening of the coexistence region relative to the monodisperse
limit.

\end{abstract}

\maketitle

\section{Introduction}

Many complex fluids, whether natural or synthetic in origin, comprise
mixtures of {\em similar} rather than {\em identical} constituents.
Examples are to be found in a host of soft matter materials. For
instance, a colloidal dispersion may contain particles which exhibit a
range of sizes, surface charge or chemical character; while many
synthetic materials contain macromolecules having a range of chain
lengths. This dependence of particle properties on one or more
continuous parameters is termed polydispersity. It affects the
performance of materials in applications ranging from foodstuffs to
polymer processing \cite{LARSON99}.

In describing polydisperse systems, it is usual to label the
polydisperse attribute by a continuous variable $\sigma$. The state of
the system is then specified by a density distribution
$\rho(\sigma)$ measuring the number density of particles of each
$\sigma$. Certain systems such as micelles, may exhibit variable
polydispersity in which the form of $\rho(\sigma)$ depends on the
prevailing chemical and thermodynamic conditions. Others such as
colloids and polymers exhibit so-called fixed polydispersity because
$\rho(\sigma)$ is set by the synthesis of the fluid. In this letter we
shall focus on the latter case.

Polydisperse fluids differ from their monodisperse counterparts in a
variety of aspects. Principal among these is the much richer character
of their phase behaviour \cite{SOLLICH02}. This richness is traceable to {\em
fractionation} effects. At phase coexistence, particles of each
$\sigma$ may partition themselves {\em unevenly} between two (or more)
coexisting `daughter' phases as long as--due to particle
conservation--the overall composition $\rho^\circ(\sigma)$ of the
`parent' phase is maintained. This partitioning alters the character of
phase diagrams. For example, the conventional liquid-gas binodal of a
monodisperse system (which connects the ends of tie-lines in a
density-temperature diagram) splits into a `cloud' and a `shadow'
curve. These give, respectively, the density at which phase coexistence
first occurs and the density of the incipient phase; the curves do not
coincide because the shadow phase in general differs in composition
from the parent. Only recently has experimental work started to elucidate
in a systematic fashion the generic consequences of fractionation for
phase coexistence properties \cite{EVANS98,FAIRHURST04,HEUKELUM03}. 

Computational solutions for dealing with polydispersity have generally
focused on the semi-grand canonical ensemble (SGCE).  Within this
framework, the instantaneous form of $\rho(\sigma)$ is permitted to
fluctuate under the control of a distribution of chemical potential
{\em differences} $\mu_d(\sigma)$, subject to a fixed overall number of
particles. Use of such an approach is attractive because it permits the
sampling of many different realizations of the ensemble of particle
sizes, thereby ameliorating finite-size effects. For the investigation
of phase coexistence, the SGCE has been combined with Gibbs-Duhem
integration \cite{BOLHUIS,KOFKE93} and Gibbs ensemble simulations
\cite{STAPLETON,KRISTOF}. However, these studies were restricted to the
case of variable polydispersity; no attempts were made to target a
specific form of $\rho(\sigma)$ or determine cloud and shadow curves.

The computational difficulties associated with tackling fixed
polydispersity are potentially quite severe: one needs to determine
that form of the chemical potential distribution $\mu(\sigma)$ for
which the ensemble averaged composition distribution $\bar\rho(\sigma)$
matches the target i.e.\ the prescribed parent $\rho^\circ(\sigma)$.
Unfortunately, the chemical potential distribution is unavailable {\em
a priori}, it being an unknown {\em functional} (i.e.\
$\mu(\sigma)=\mu[\rho^\circ(\sigma)]$), of the parent. Recently
however, techniques have been developed that efficiently overcome this
difficulty within the framework of a full grand canonical ensemble
(GCE) \cite{WILDING02,ESCOBEDO01,WILDING03}. The latter is particular
well suited to the study of fluid phase transitions due to the
fluctuating overall particle number. In this letter we demonstrate that
when combined with extended sampling methods, the new techniques
facilitate the detailed and efficient study of phase behaviour in
fluids of fixed polydispersity.

\section{Computational aspects}
\label{sec:method}

The grand canonical ensemble Monte Carlo algorithm we employ has been
described in ref.~\cite{WILDING02} and invokes four types of operation:
particle displacements, deletions, insertions, and resizing. The
polydisperse attribute $\sigma$ is itself represented as a strictly
continuous variable, subject to some upper bound $\sigma_c$. However,
observables such as the instantaneous composition distribution
$\rho(\sigma)$ are accumulated in the form of a histogram by
discretising the $\sigma$ domain into a prescribed number of bins. This
discretisation also applies to the chemical potential distribution
$\mu(\sigma)$, i.e.\ all particles whose $\sigma$ values is encompassed
by the same bin are subject to an identical chemical potential.

The form of the ensemble averaged composition distribution
$\bar\rho(\sigma)$ is controlled by $\mu(\sigma)$, via its role in the
acceptance probabilities for particle transfers and resizing moves. For
fixed polydispersity one wishes to match $\bar\rho(\sigma)$ to the
desired parent distribution. The latter can be written as

\begin{equation}
\rho^\circ(\sigma)=n_0 f(\sigma),
\label{eq:parent}
\end{equation}
where $n_0=N/V$ is the overall particle number density, while
$f(\sigma)$ is a prescribed normalized shape function. Since
$\rho^\circ(\sigma)$ may vary only in terms of its scale $n_0$, the
system is constrained to traverse a {\em dilution line} in the full
phase space of possible compositions. The task is then to determine, as
a function of temperature, the form of $\mu(\sigma)$ along the dilution
line. More specifically, we seek the intersection of the dilution line
with a coexistence region. Recently developed simulation techniques
facilitate this, as we now summarize.

The non-equilibrium potential refinement (NEPR) scheme \cite{WILDING03}
permits the efficient iterative determination of
$\mu[\rho^\circ(\sigma),T]$, from a single simulation run, and without
the need for an initial guess of its form. To achieve this, the method
continually updates $\mu(\sigma)$ in such as way as to minimize the
deviation of the instantaneous density distribution $\rho(\sigma)$
from the target form (i.e.\ the parent). However, tuning $\mu(\sigma)$
in this manner clearly violates detailed balance. To counter this,
successive iterations reduce the degree of modification applied to
$\mu(\sigma)$, thereby driving the system towards equilibrium and ultimately
yielding the equilibrium form of $\mu[\rho^\circ(\sigma),T]$.

For the purpose of exploring phase diagrams, Histogram Extrapolation
(HE) techniques have proved invaluable \cite{HR}. In the present
context, their use permits histogram of observables accumulated at one
$\mu(\sigma)$ to be reweighted to estimate observables at some other
$\mu(\sigma)$. In ref.~\cite{WILDING02} we have shown how HE can be
combined with a minimization scheme, to track a dilution line in a
stepwise fashion. We shall deploy this approach again in the present
study.

Simulation studies of phase coexistence present distinctive challenges.
Principal among these is the large free energy (surface tension)
barrier separating the coexisting phases. In order to accurately locate
coexistence points, a sampling scheme must be utilized which enables
this barrier to be surmounted \cite{BRUCE03}. One such scheme is
multicanonical preweighting, which utilizes a weight function in the MC
acceptance probabilities, in order to encourage the simulation to
sample the interfacial configurations of low probability \cite{BERG92}.
At a given coexistence state point, the requisite weight function takes
the form of an approximation to the inverse of the distribution of the
fluctuating number density, $p(n)$, with $n=N/V$. While specialized techniques
allow determination of $p(n)$ from scratch, in situations where
one wishes to track a fluid-fluid phase boundary, prior determination
of a weight function is unnecessary provided one commences from the
vicinity of the critical point where the barrier to inter-phase
crossings is small. Data accumulated here can be used (together with
HE) to provide estimates of suitable multicanonical weight functions at
lower temperatures \cite{WILDING95} where the barrier height is
greater.

\section{Model}

The techniques outlined above have been deployed to obtain the
liquid-vapor coexistence properties of a fluid of particles interacting
via a pairwise potential of the Lennard-Jones (LJ) form:

\begin{equation}
U(r_{ij}, \sigma_{ij})=4\epsilon\left[\left(\frac{\sigma_{ij}}{r_{ij}} \right)^{12}-\left(\frac{\sigma_{ij}}{r_{ij}} \right)^6\right]\:,
\end{equation}
with $\sigma_{ij}=(\sigma_i+\sigma_j)/2$. A cutoff was applied to this
potential for particle separations $r_{ij}>2.5\sigma_{ij}$. 

The polydispersity enters solely through the distribution of diameters
$\sigma_i$. Our algorithm finds that $\mu(\sigma)$ for which the
diameters are distributed according to $f(\sigma)$ given a choice for
$n_0$ (cf. eq.~\ref{eq:parent}). We have assigned $f(\sigma)$ the Schulz form:

\begin{equation}
f(\sigma)=\frac{1}{z!}\left(\frac{z+1}{\bar{\sigma}}\right)^{z+1}\sigma^z\exp\left[-\left(\frac{z+1}{\bar{\sigma}}\right)\sigma\right]\:.
\label{eq:schulz}
\end{equation}
Here $\bar\sigma\equiv 1$ is the average particle diameter, while $z$ is a
width parameter, the value of which was set to $z=5$, corresponding to
a high degree of polydispersity (standard deviation of $f(\sigma)$), $\delta=40\%$.
Additionally, for convenience, $f(\sigma)$ was truncated at
$\sigma_c=3.0$. Histograms of observables were formed by discretising the permitted
range $0\le \sigma \le \sigma_c$ into $120$ bins.

\section{Simulation strategy and results}
\label{sec:results}

The strategy employed for mapping the liquid-vapor coexistence curve of
our model was as follows. In order to bootstrap the dilution line
tracking procedure, the NEPR method \cite{WILDING03} was employed to
determine $\mu(\sigma)$ for a gas phase state point on the dilution
line at a moderately low temperature.  Starting from this point, the dilution
line was then followed towards increasing density (with the aid of HE)
until the gas spontaneously liquefied. Having estimated the location of
a coexistence state point in this manner, the temperature was increased
in steps (whilst remaining on the dilution line) until the density
difference between the gas and the spontaneously formed liquid
vanished, signalling the proximity of the critical point. Finite-size
scaling methods \cite{WILDING95} were then used to home in on the
critical parameters.

\begin{figure}[h]
\includegraphics[width=8.0cm,clip=true]{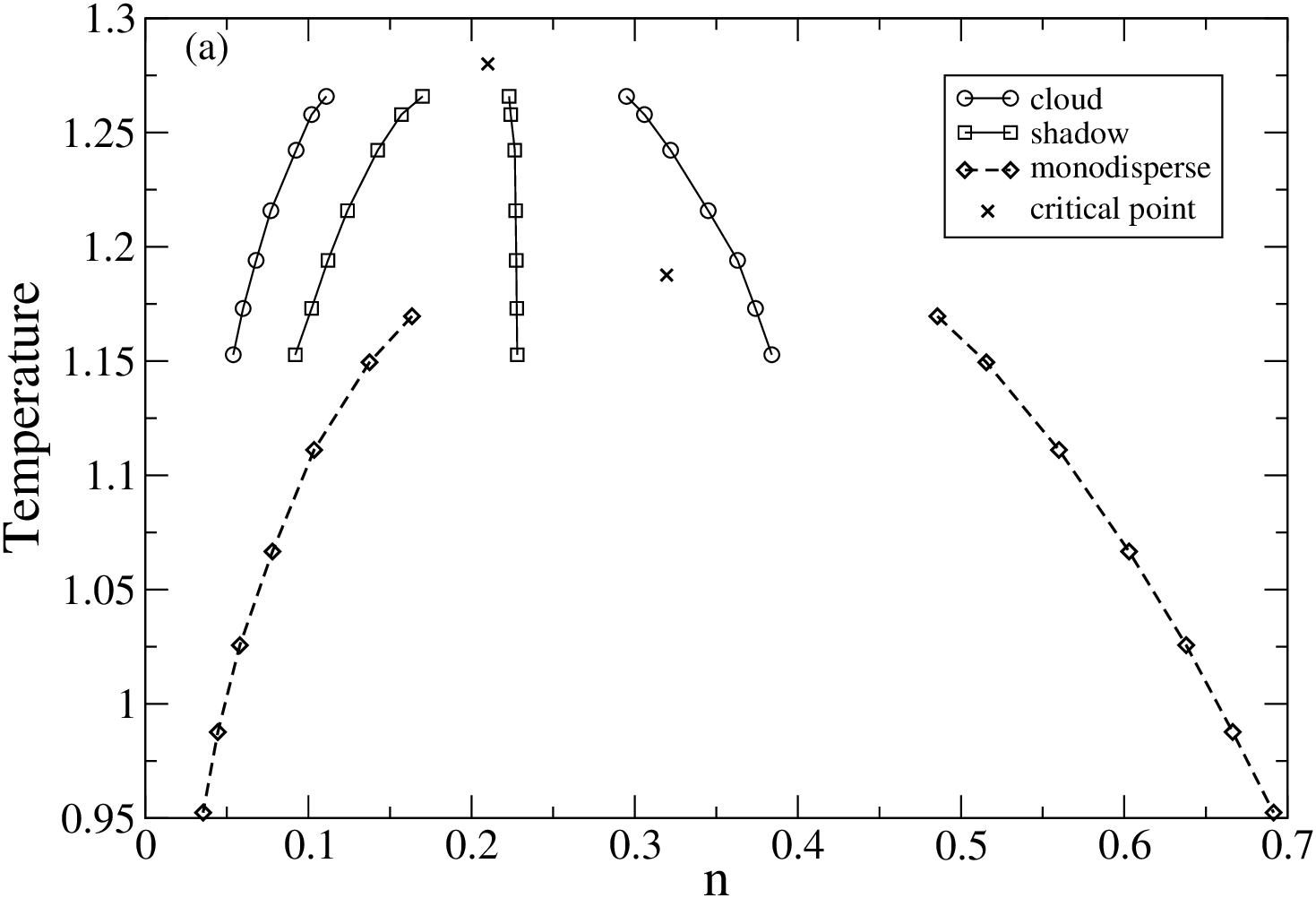}
\includegraphics[width=8.0cm,clip=true]{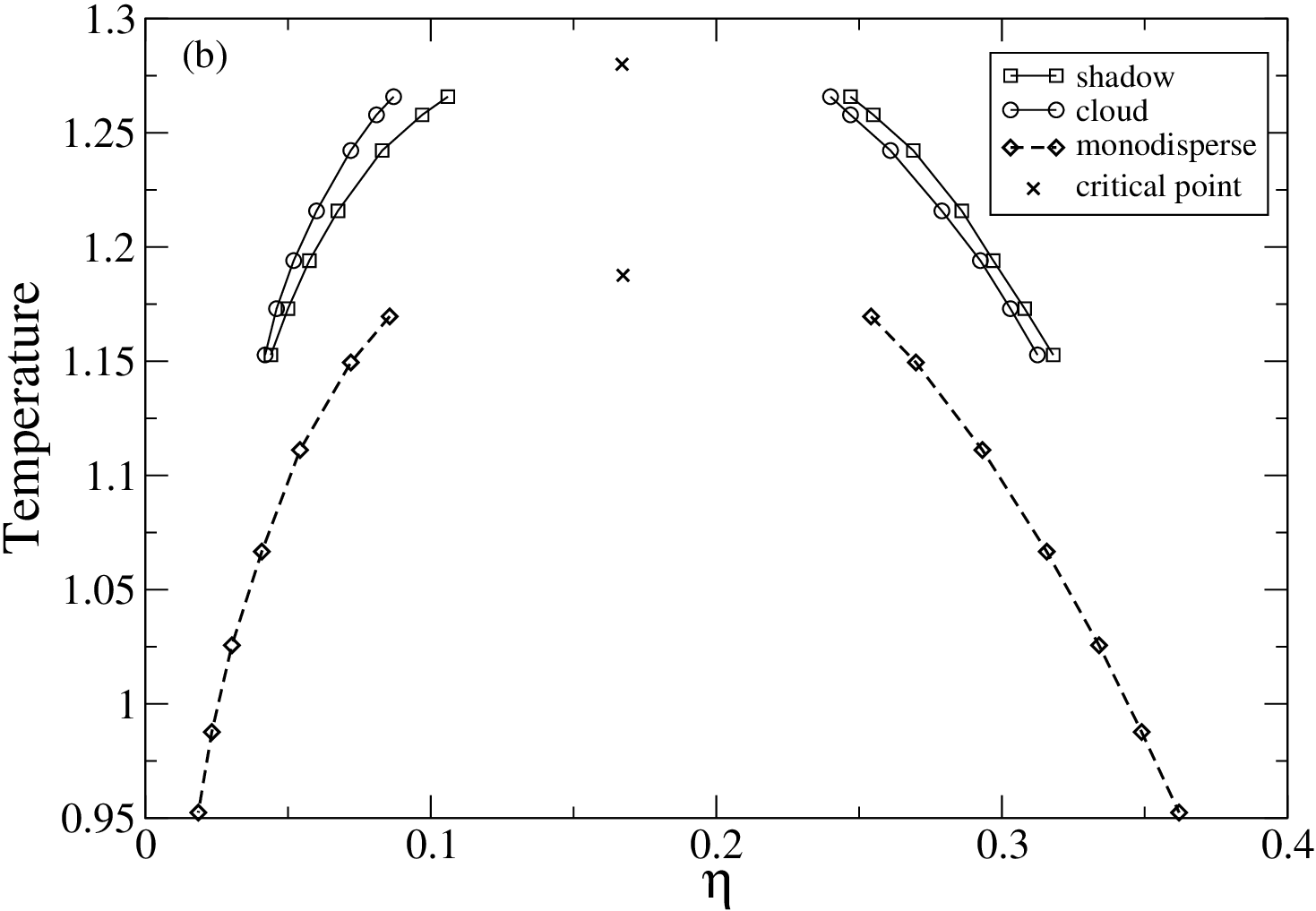}
\caption{The measured phase diagram of the
size-disperse LJ fluid. {\bf (a)} The $n-T$
representation. {\bf (b)} The $\eta-T$ representation. In both cases, the phase diagram of the monodisperse
limit (broken line) is shown for comparison. Critical points (determined by
finite-size scaling) are shown as crosses. Statistical errors do not
exceed the symbol sizes.}
\label{fig:densities}
\end{figure}

Having located the critical point, a detailed mapping of the cloud and
shadow curves was performed for a large simulation box of volume
$V=11390\bar\sigma^3$. Attention was focused on the distribution of the
fluctuating overall number density, $p(n)$. The gas phase cloud point
(incipient liquid phase) corresponds to the situation where $p(n)$
is bimodal, but with vanishingly small weight in the liquid peak. Under
these conditions, the position of the low density gas peak provides an
estimate of the gas phase cloud density, while that of the liquid peak
gives the gas phase shadow density. The converse is true for the
liquid phase cloud point and its shadow. Determining the cloud and
shadow points as a function of temperature yields the cloud and shadow
curves. We have tracked the gas and liquid cloud curves (and their
shadows) in a stepwise fashion downwards in temperature from the
critical point. Histogram extrapolation was employed to negotiate each
temperature step, yielding estimates for both the form of $\mu(\sigma)$
on the cloud curve at the next temperature, and the requisite
multicanonical weight function. 

The resulting phase diagram is shown in fig.~\ref{fig:densities}(a),
together with that of the monodisperse LJ fluid, determined in an
earlier study \cite{WILDING95}. It should be pointed out that while the
positions of the peaks in $p(n)$ provide an accurate estimate of
cloud and shadow points at low temperatures, this breaks down near the
critical point due to finite-size effects \cite{WILDING95}. Thus a
naive extrapolation of our curves to their intersection point will tend
to overestimate the critical temperature. However, our independent
determination of the critical point using finite-size scaling methods
(as indicated in fig.~\ref{fig:densities}(a)) is considerably more accurate. 

The results of fig.~\ref{fig:densities}(a) show a stark separation of
the cloud and shadow curves in the $n-T$ plane. Furthermore, the
whole phase diagram is considerably shifted with respect to that of the
monodisperse fluid. Specifically, one observes that the critical point
occurs at a considerably higher temperature than in the monodisperse
limit. This particular finding contrasts with that of a previous
theoretical study of a size-disperse van-der Waals fluid
\cite{BELLIER00}, which predicts a {\em suppression} of the critical
temperature with respect to the monodisperse limit.

\begin{figure}[h]
\includegraphics[width=8.0cm,clip=true]{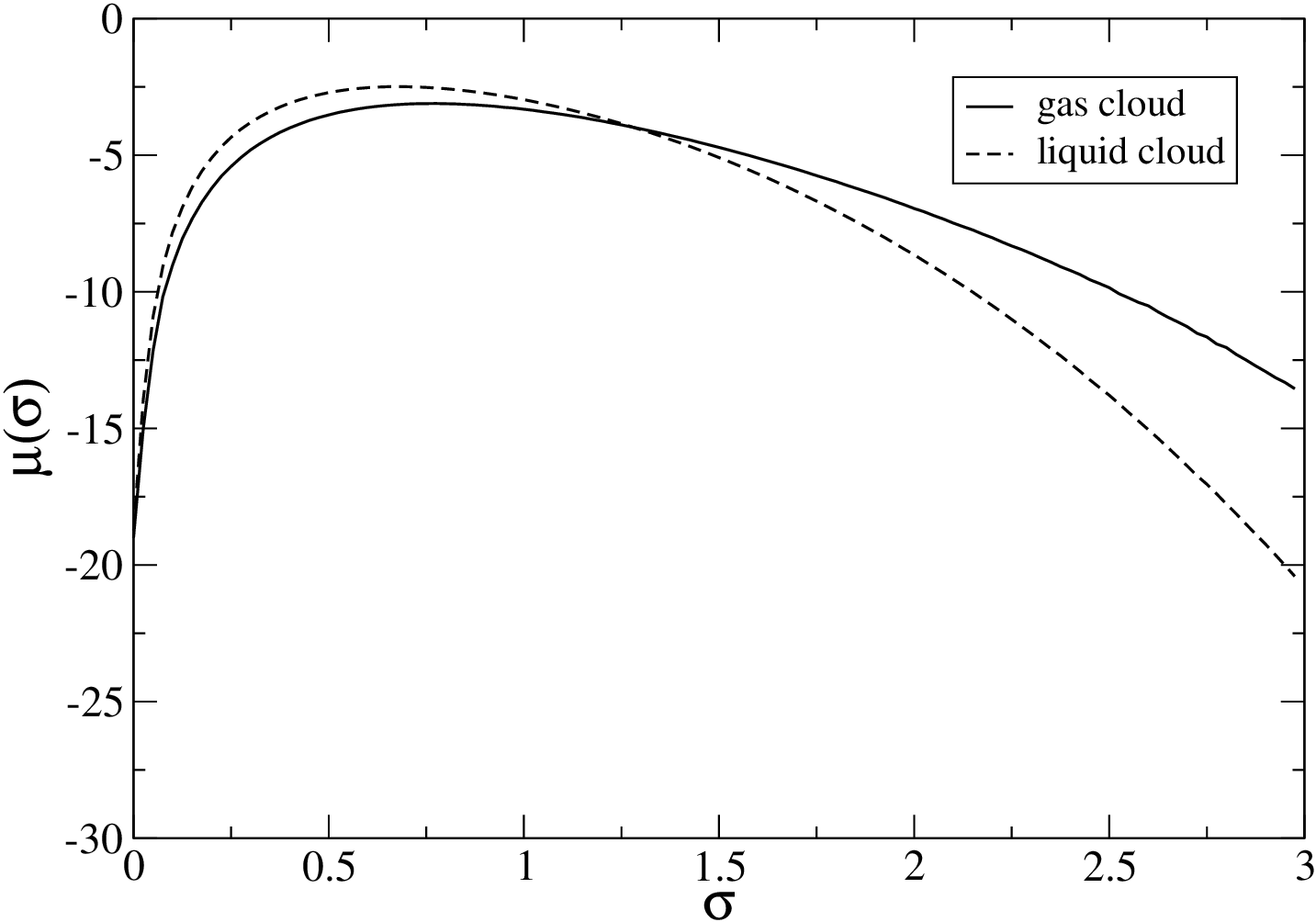}
\caption{The measured form of the chemical potential distribution
$\mu(\sigma)$ at the gas and liquid phase cloud points for $T=0.91T_c$. Statistical errors do not
exceed the symbol sizes.}
\label{fig:mu}
\end{figure}

The standard order of cloud and shadow curves with increasing density is
cloud-shadow-cloud-shadow. By contrast, the  pattern apparent in
fig.~\ref{fig:densities}(a) is cloud-shadow-shadow-cloud.
Interestingly, however, the order reverts to the standard pattern if
one plots the data in terms of the volume fraction $\eta=(\pi/6)\int
d\sigma \sigma^3\rho(\sigma)$, rather than the number density, as shown
in fig.~\ref{fig:densities}(b). Moreover, one sees that in the $\eta-T$ representation
the differences between cloud and shadow phase properties become much
less pronounced. In particular, while the critical number density of
the polydisperse fluid is considerably less than its value in the
monodisperse limit, the critical volume fraction for the mono- and
polydisperse fluid agree to within error. However, irrespective of the
choice of data representation, we observe that for our model the
critical point occurs very close to the top of the coexistence curve.
No clear evidence was discernible for distinct cloud and shadow points,
at or above $T_c$.

Notwithstanding these intriguing findings, not all differences between
gas and liquid cloud points at a given temperature can be camouflaged
by a simple change of variable. At temperatures significantly below
criticality, we observe dramatic broadening of the coexistence curve in
the space of $\mu(\sigma)$. This is shown in fig.~\ref{fig:mu} which
presents the form of $\mu(\sigma)$ at the respective cloud points for
the lowest temperature studied, $T=0.91T_c$. Such broadening does not
occur in monodisperse systems (coexistence occurs at a single value of
the chemical potential, not a range of values). The effect is
surprisingly large, even given the high degree of polydispersity of the
parent ($\delta=40\%$). Indeed, in simulation terms, the respective
cloud points are so far separated in phase space that to connect them
directly (via a route crossing the phase boundary) required a dozen
overlapping simulations--twice as many as were required to connect the
cloud point to the critical point at this temperature. We remark in
passing that similar aspects of coexistence curve broadening have
recently been analyzed within the context of Landau theory by Rascon
and Cates \cite{RASCON03}. 

\begin{figure}[h]
\includegraphics[width=7.0cm,clip=true]{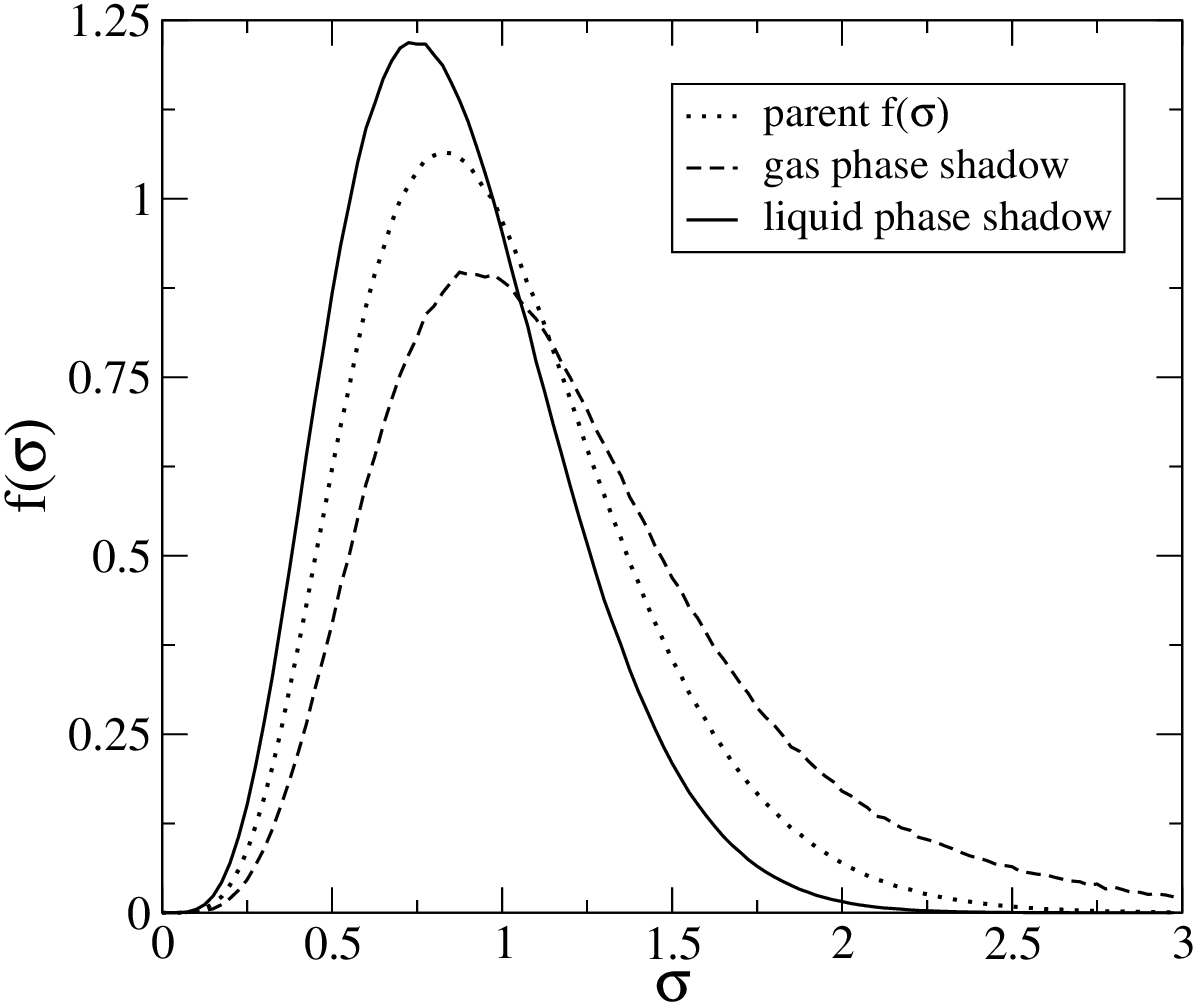}

\caption{The normalized form of the particle size distribution at the
gas and liquid phase shadow points at $T=0.91T_c$. The average particle
diameter in the gas phase shadow is $\bar\sigma=1.167(3)$, while that in the
liquid-phase shadow is $\bar\sigma=0.863(2)$. Also shown for
comparison is the parent shape function $f(\sigma)$, corresponding to
the respective cloud phase distributions.}
\label{fig:fraction}
\end{figure}

Finally in this section we present the normalized daughter phase
distributions at the gas and liquid cloud points for $T=0.91T_c$. The
data show that at the gas phase cloud point, larger particles preferentially 
occupy the shadow phase. Conversely at the liquid phase cloud point,
there is a predominance of smaller particles in the shadow phase.
Clearly the scale of these fractionation effects is considerable: the
polydispersity of the gas phase shadow at this temperature is close to
$50\%$, while that of the liquid phase shadow is $\approx 33\%$, to be
compared with a parent polydispersity of $40\%$. 

\section{Discussion and conclusions}
\label{sec:conc}

In summary, we have demonstrated how extended sampling grand canonical
simulations can be combined with histogram extrapolation methods and a
new non-equilibrium potential refinement scheme to accurately determine
the phase behaviour of a polydisperse fluid. The results show that in
contrast to existing theoretical predictions, the critical temperature
of the polydisperse system exceeds that of its monodisperse
counterpart. As regards the sub-critical region, we find that the
relative order of cloud and shadow curves changes depending on whether
the data is represented in terms of the overall number density or the
volume fraction. Additionally, we observe considerable
polydispersity-induced broadening of the coexistence region: at a given
temperature, the cloud points of the respective phases (which coincide
in a monodisperse system) are widely separated in terms of their
chemical potential distributions. The scale of this effect is mirrored
in the disparate forms of the shadow phase daughter distributions.

In a future publication \cite{WILDING04}, we will present further
simulation results for the magnitude of critical point shifts as
a function of the width of the governing size distribution $f(\sigma)$.
These results will be compared with those of a moment based theoretical
analysis \cite{SOLLICH98,SOLLICH01} of an improved model free energy
for the size disperse van der Waals fluid. The latter correctly
captures the sign of polydispersity-induced critical point shifts and
provides insights into the deficiencies of previous approaches.

\acknowledgments

The authors acknowledge support of the EPSRC, grant numbers GR/S59208/01
and GR/R52121/01.

\end{document}